\documentclass[preprint,5p,12pt,twocolumn]{elsarticle}

\usepackage{amssymb}
\usepackage{amsmath}

\usepackage{tikz}
\usepackage{subcaption}
\usepackage{siunitx}
\usepackage{xcolor}
\usepackage{hyperref}
\usepackage{xurl}
\hypersetup{breaklinks=true}
\usepackage[hyphenbreaks]{breakurl}
\hypersetup{ colorlinks, 
	linkcolor={red!50!black}, 
	citecolor={blue!80!black}, 
	urlcolor={blue!80!black} }
\usepackage{cuted} 

\graphicspath{{figs/}}

\definecolor{airforceblue}{rgb}{0.36, 0.54, 0.66}
\definecolor{ballblue}{rgb}{0.13, 0.67, 0.8}
\definecolor{cobalt}{rgb}{0.0, 0.28, 0.67}

\journal{Solar Energy Materials and Solar Cells}

\begin{document}

\begin{frontmatter}

\title{Radiator Tailoring for Enhanced Performance in InAs-Based Near-Field Thermophotovoltaics}

\author[1]{Mathieu Giroux\corref{cor1}}
\address[1]{Department of Mechanical Engineering, University of Ottawa, Ottawa K1N 6N5, Ontario, Canada}
\ead{mgiro027@uottawa.ca}
\cortext[cor1]{Corresponding author.}

\author[2]{Sean Molesky}
\address[2]{Department of Engineering Physics, Polytechnique Montreal, Montreal H3T 1J4, Quebec, Canada}

\author[1,3,4]{Raphael St-Gelais}
\address[3]{Department of Physics, University of Ottawa, Ottawa K1N 6N5, Ontario, Canada}
\address[4]{Nexus for Quantum Technologies, University of Ottawa, Ottawa K1N 6N5, Ontario, Canada}

\author[3,4]{Jacob J. Krich}


\begin{abstract}
Near-field thermophotovoltaics (NFTPV) systems have significant potential for waste heat recovery applications, with both high theoretical efficiency and power density, up to 40\% and \SI{11}{\watt/\centi\meter\squared} at 900 K. Yet experimental demonstrations have only achieved up to 14\% efficiency and modest power densities (i.e., \SI{0.75}{\watt/\centi\meter\squared}). While experiments have recently started to focus on photovoltaic (PV) cells custom-made for NFTPV, most work still relies on conventional doped silicon radiators. In this work, we design an optimized NFTPV radiator for an indium arsenide-based system and, in the process, investigate models for the permittivity of InAs in the context of NFTPV. Based on existing measurements of InAs absorption, we find that the traditional Drude model overestimates free carrier absorption in InAs. We replace the Drude portion of the InAs dielectric function with a revised model derived from ionized impurity scattering. Using this revised model, we maximize the spectral efficiency and power density of a NFTPV system by optimizing the spectral coupling between a radiator and an InAs PV cell. We find that when the radiator and the PV cell are both made of InAs, a nearly threefold improvement of spectral efficiency is possible compared to a traditional silicon radiator with the same InAs cell. This enhancement reduces subgap thermal transfer while maintaining power output.   
\end{abstract}


\end{frontmatter}



\section{Introduction}
\label{sec:intro}

\textbf{}Waste heat above 600 K represents 10\% of the global energy consumption \cite{Forman_2016}. Harvesting this energy would significantly enhance energy-use efficiency and reduce carbon emissions. For 600-900 K heat sources, thermoelectric generators (TEGs) are the most-widely used solid-state technology \cite{Tedah_2019}. While TEGs have experimentally achieved high power densities up to \SI{22}{\watt/\centi\meter\squared} at 868 K \cite{He_2016}, conduction losses limit achievable efficiencies \cite{Datas_2019}, which are typically less than 10\% \cite{Tedah_2019}. Thermophotovoltaic (TPV) systems present an alternative, potentially efficient, solid-state solution, with theoretical efficiencies up to 45\% at 900 K \cite{Tedah_2019}, but are limited to low power densities, below \SI{0.2}{\watt/\centi\meter\squared} at 900 K \cite{Tedah_2019}. Near-field thermophotovoltaics (NFTPV) systems have significant potential for waste heat recovery applications \cite{Laroche_2006,Zhao_2017,Molesky_2015,Ilic_2012,St-Gelais_2017}, with both high theoretical efficiency and power density, up to 40\% and \SI{11}{\watt/\centi\meter\squared} at 900 K \cite{Zhao_2017}. This technology exploits the drastic enhancement of the thermal radiation between two bodies at subwavelength distances due to evanescent radiative thermal coupling. When a cold photovoltaic (PV) cell is positioned in such close proximity to a hot radiator, tremendous amounts of heat transfer from the hot object to the PV cell, where it can be converted to electricity. \par

Despite such potential, experimental demonstrations have only achieved up to 14\% efficiency and \SI{0.75}{\watt/\centi\meter\squared} power density \cite{Fiorino_2018,Bhatt_2020,Lucchesi_2021,Mittapally_2021}. One of the main challenges in the development of NFTPV technology is the need for specialized narrow-bandgap PV cells. Early demonstrations \cite{Fiorino_2018,Bhatt_2020} relied on basic PV cells fabricated in house \cite{Bhatt_2020} or on commercially available photodetectors \cite{Fiorino_2018}. The use of PV cells unoptimized for near-field operations resulted in modest performance, with conversion efficiencies below 1\% and less than 1 nW of generated power. Recently, PV cells optimized for NFTPV operation \cite{Lucchesi_2021,Mittapally_2021} were reported, leading to higher efficiencies and power densities, up to 14\% and \SI{0.75}{\watt/\centi\meter\squared}, which are still far from the predicted limits \cite{Laroche_2006,Zhao_2017,Molesky_2015,Ilic_2012,St-Gelais_2017}. \par

One reason for the gap between high theoretical performances and modest experimental results is that most experimental work still relies on conventional doped Si radiators \cite{Fiorino_2018,Mittapally_2021,Inoue_2019}. A few theoretical works reported optimized radiator designs for NFTPV applications \cite{Ilic_2012,St-Gelais_2017}. These studies relied on plasmonic materials, such as indium-tin-oxide (ITO), which exhibit surface polariton resonances in the infrared region, enhancing the radiative heat flux in a spectral distribution located just above the bandgap of the PV cell. While these materials greatly improve the theoretical performance of NFTPV systems, in practice their optical properties are contingent upon factors such as film annealing and deposition conditions \cite{Alam_2002}, making the experimental implementation of these proposed films quite challenging. Here, with eye towards near-term large-scale implementation, we focus on the more practical and reproducible possibilities offered by crystalline materials. \par

For the temperature range of waste heat recovery applications ($<$ 1000 K \cite{Fiorino_2018}), InAs is a promising material \cite{Zhao_2017,Forcade_2022,Selvidge_2024}, with a bandgap (0.35 eV) well aligned to the Planck spectrum. To predict the properties of InAs in a NFTPV system, an accurate dielectric model is imperative to correctly model the radiative heat transfer spectrum. For this purpose, most dielectric models employ a Drude form for the free-carrier absorption. \par

We have found that this choice leads to a consistent overestimation of free carrier absorption, both above and below the bandgap, especially for highly doped InAs. Here, we correct this overestimation by substituting the Drude response for a revised model, derived from ionized impurity scattering \cite{Baltz_1972}, and, using this refinement, design an optimized radiator for an InAs-based NFTPV system. \par

We compare the performance of a doped silicon radiator to that of a doped InAs radiator, both paired with an InAs-based photovoltaic cell designed specifically for NFTPV \cite{Forcade_2022}. We find that, when the doping levels and thicknesses of the radiators are chosen optimally, the power output from both radiators is nearly identical, but the spectral efficiency is higher with an InAs radiator. Namely the use of a silicon radiator leads to increased subgap heat transfer, due to an additional subgap resonance. In addition to improving the heat transfer spectrum, reducing such subgap heat transfer lowers cooling requirements and is therefore highly desirable.\par

\section{Results and Discussion}
\label{sec:results}

\subsection{Dielectric Function}
\label{subsec:dielectric_function}

Calculating radiative heat transfer requires knowledge of the dielectric functions of all materials used (in our case Si and InAs) from low energy up to several times the material bandgap. For Si, we use a dielectric model containing lattice and interband contributions from Ref.~\citenum{Lee_2005} combined with the Drude free-carrier model from Ref.~\citenum{Fu_2006}, while incorporating the mobility, decay rates and donor and acceptor contributions from Ref.~\citenum{Basu_2010}. All of these models have been validated against experimental data in Refs.~\citenum{Lee_2005,Fu_2006}, and~\citenum{Basu_2010}. For InAs, we employ the comprehensive dielectric model assembled in Ref.~\citenum{Milovich_2020} with the updated parameters from Ref.~\citenum{Forcade_2022}. This dielectric model includes a Drude-Lorentz model to account for both lattice and free carrier contributions, as well as an interband absorption model with non-parabolicity corrections and the Moss-Burstein shift. Note that, unlike Ref.~\citenum{Milovich_2020}, we employ an approximation to the Kramers-Kronig relations to improve computational speed (see \ref{Appendix_A} for full discussion). A Python implementation of these dielectric models can be found in Ref.~\citenum{Giroux_GitHub_2024}.\par

\begin{figure}[!htb]
	\begin{subfigure}{\columnwidth}
		\captionsetup{justification=raggedright, singlelinecheck=false, position=top} 
		\includegraphics[width=\columnwidth]{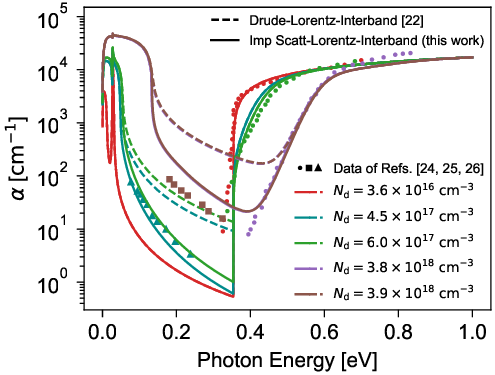}
		\vspace{-71mm}
		\caption{}\label{fig_1:sub1} 
		\vspace{65.5mm}
	\end{subfigure}
	\begin{subfigure}{\columnwidth}
		\captionsetup{justification=raggedright, singlelinecheck=false, position=top} 
		\includegraphics[width=\columnwidth]{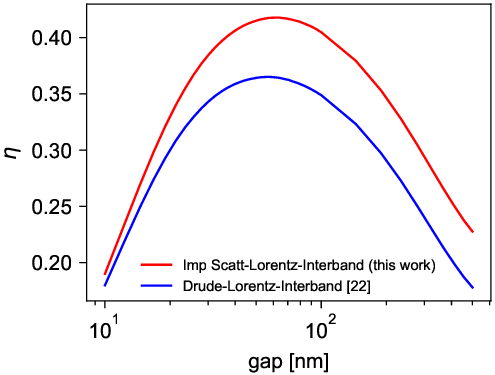}
		\vspace{-71.5mm}
		\caption{}\label{fig_1:sub2} 
		\vspace{64.75mm}
	\end{subfigure}
	\caption{(\subref{fig_1:sub1}) InAs absorption coefficient from the dielectric model of Ref.~\citenum{Milovich_2020} (dashed lines), which combines a Drude-Lorentz model with an interband model (Drude-Lorentz-Interband), compared to data from n-doped InAs (markers). Our revised model, which replaces the Drude model with an ionized impurity scattering model (Imp Scatt-Lorentz-Interband), significantly improves the agreement with experimental data above and below the bandgap, especially at high doping. Circles from Ref.~\citenum{Dixon_1961}, squares from Ref.~\citenum{Dixon_1960}, and triangles from Ref.~\citenum{Culpepper_1968}. (\subref{fig_1:sub2}) Computed spectral efficiency, with the Drude-Lorentz-Interband model (blue) and with our proposed model (red), as a function of the gap for an InAs-based NFTPV system using an InAs radiator. The Drude model underestimates the spectral efficiency by up to 22\% because it overestimates below-bandgap free carrier absorption.}\label{fig1}
\end{figure}

Based on existing measurements of InAs absorption \cite{Dixon_1961,Dixon_1960,Culpepper_1968}, we find that the traditional Drude model, proposed for example in Ref.~\citenum{Milovich_2020}, overestimates free carrier absorption in InAs. Fig.~\ref{fig_1:sub1} compares the InAs absorption coefficient from the dielectric model described in Ref.~\citenum{Milovich_2020} to experimental n-doped InAs absorption data at different donor concentrations $N_\mathrm{d}$ taken from Refs.~\citenum{Dixon_1961,Dixon_1960}, and~\citenum{Culpepper_1968}. The comparison suggests that the Drude contribution results in a significant overestimation of free carrier absorption at energies near the bandgap. This discrepancy becomes more noticeable at higher doping levels, leading to an overestimation of absorption above the bandgap. These differences cause inaccurate model predictions of NFTPV performances as above-bandgap absorption originating from free carrier absorption does not generate collectible electron-hole pairs. Additionally, parasitic subgap heating has a considerable effect on the overall performance of the device. \par

We address this overestimation by incorporating an alternative model, described in Ref.~\citenum{Baltz_1972}, derived from ionized impurity scattering. The corrective model is only valid when the Fermi level $E_\mathrm{F}$ measured from the conduction band minimum is larger than the thermal energy and the frequency $\omega$ exceeds the plasma frequency $\omega_\mathrm{p}$, given by

\begin{equation}
\omega_\mathrm{p}=\sqrt{\frac{Ne^2}{\varepsilon_0\varepsilon_\infty m^*_\mathrm{e}}},
\end{equation}\label{eq:w_p}

\noindent where $N$ is the carrier density, $e$ is the elementary charge, $\varepsilon_0$ is the vacuum permittivity, $\varepsilon_\infty$ is the high-frequency dielectric constant, and $m^*_\mathrm{e}$ is the electron conduction band effective mass. Therefore, we still employ the Drude model when $\omega<\omega_\mathrm{p}$ or when $E_\mathrm{F}$ is below the thermal energy. For highly n-doped InAs, when $\omega>\omega_\mathrm{p}$, the corrective model substitutes the imaginary component of the Drude model $\varepsilon^\mathrm{\prime\prime}_\mathrm{FC}$ with

\begin{strip}
	\begin{align}
	\varepsilon^\mathrm{\prime\prime}_\mathrm{FC}(\omega) = &A \zeta^{-4} \int_{\left(1-\zeta\right)\theta\left(1-\zeta\right)}^1 \left[\frac{1}{2}\ln\frac{\left(\sqrt{X+\zeta}+\sqrt{X}\right)^2+X_{\mathrm{TF}}}{1\left(\sqrt{X+\zeta}-\sqrt{X}\right)^2+X_{\mathrm{TF}}} \right. \nonumber\\
	&- \left. \frac{2X_{\mathrm{TF}}\sqrt{X\left(X+\zeta\right)}}{\left[\left(\sqrt{X+\zeta}+\sqrt{X}\right)^2+X_{\mathrm{TF}}\right]\left[\left(\sqrt{X+\zeta}-\sqrt{X}\right)^2+X_{\mathrm{TF}}\right]}\right]dX \label{eq:Baltz_model}
	\end{align}
\end{strip}

\noindent where $\zeta=\frac{\hbar\omega}{E_\mathrm{F}}$, $X_\mathrm{TF}=\left(\frac{q_\mathrm{TF}}{k_\mathrm{F}}\right)^2$, $q_\mathrm{TF}=\sqrt{\frac{3Ne^2}{2\varepsilon_0\varepsilon_\infty E_\mathrm{F}}}$ is the Thomas-Fermi screening wavevector, $A=\frac{1}{12\pi^3}\frac{e^2\gamma k^4_\mathrm{F}}{\varepsilon_0 E^3_\mathrm{F}}$, $k_\mathrm{F}=\frac{\sqrt{2m^*_\mathrm{e} E_\mathrm{F}}}{\hbar}$ is the Fermi wavevector, $\gamma=\left[\frac{Ze^2}{\varepsilon_0\varepsilon_\infty k^2_\mathrm{F}}\right]^2$, and $\theta(x)$ is the Heaviside step function. Here, $\hbar$ is the reduced Planck constant, $Z$ is the charge number of the impurity, and $R$ denotes the number of impurities with charge $Z\cdot e$. With this new form for $\varepsilon^\mathrm{\prime\prime}_\mathrm{FC}(\omega)$, we can use the Kramers-Kronig relations to find the corresponding modification to the real part of the dielectric $\varepsilon^\mathrm{\prime}_\mathrm{FC}(\omega)$. We have confirmed that simply using $\varepsilon^\mathrm{\prime}_\mathrm{FC}$ from the original Drude model introduces less than 1\% error in $\varepsilon^\mathrm{\prime}_\mathrm{FC}$ in a highly doped test case. \par

The proposed model more adequately represents the free carrier contribution and results in a good agreement with the experimental data at doping levels below $10^{18} \, \SI{}{\per\centi\meter\cubed}$ (see Fig.~\ref{fig1}). While this is a significant improvement, the model could be further refined as we continue to observe approximately a twofold overestimation of the free carrier contribution at high doping levels (e.g., $N_\mathrm{d}=3.8\times10^{18} \, \SI{}{\per\centi\meter\cubed}$ near 0.4 eV). \par

In Fig.~\ref{fig_1:sub2}, we compare the spectral efficiency of a NFTPV system using an optimized InAs radiator computed with a Drude model to that computed with our proposed model; details on the optimization and calculation of spectral efficiency are in the next section. We find that the standard Drude model results in up to a 22\% underestimation of the spectral efficiency due to the overestimation of below bandgap absorption.\par

\subsection{Radiator Optimization}
\label{subsec:radiator_optimization}

\subsubsection{InAs as a Radiator}
\label{subsubsec:InAs_as_radiator}

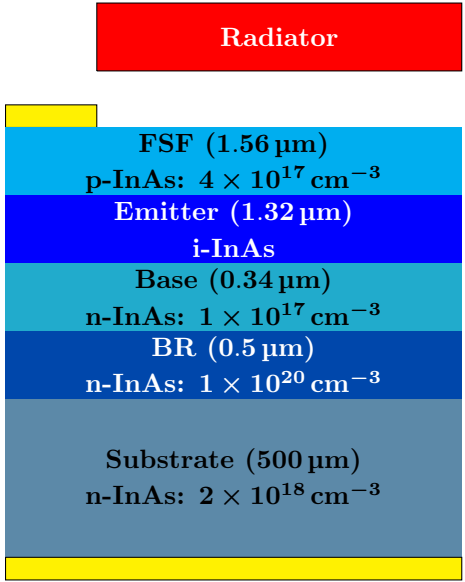
\begin{figure}[!htb]
	\centering
	\begin{tikzpicture}[scale=0.6]

		\fill [red] (2,11.25)--(10,11.25)--(10,12.75)--(2,12.75)--(2,11.25);
		\draw (2,11.25)--(10,11.25)--(10,12.75)--(2,12.75)--(2,11.25);
		\node[font=\fontsize{10}{60}\selectfont,white] at (6,12) {\textbf{Radiator}};
		
		\fill [yellow] (0,10)--(2,10)--(2,10.5)--(0,10.5)--(0,10);
		\draw (0,10)--(2,10)--(2,10.5)--(0,10.5)--(0,10);
		\fill [cyan] (0,10)--(10,10)--(10,8.5)--(0,8.5)--(0,10);
		\node[font=\fontsize{10}{60}\selectfont] at (5,9.6) {\textbf{FSF ($\mathbf{1.56} \, \textbf{\textmu m}$)}};
		\node[font=\fontsize{10}{60}\selectfont] at (5,8.85) {\textbf{p-InAs: $\mathbf{4} \boldsymbol{\times} \mathbf{10^{17}} \, \boldsymbol{\mathrm{cm^{-3}}}$}};
		\fill [blue] (0,8.5)--(10,8.5)--(10,7)--(0,7)--(0,8.5);
		\node[font=\fontsize{10}{60}\selectfont,white] at (5,8.1) {\textbf{Emitter ($\mathbf{1.32} \, \textbf{\textmu m}$)}};
		\node[font=\fontsize{10}{60}\selectfont,white] at (5,7.35) {\textbf{i-InAs}};
		\fill [ballblue] (0,7)--(10,7)--(10,5.5)--(0,5.5)--(0,7);
		\node[font=\fontsize{10}{60}\selectfont] at (5,6.6) {\textbf{Base ($\mathbf{0.34} \, \textbf{\textmu m}$)}};
		\node[font=\fontsize{10}{60}\selectfont] at (5,5.85) {\textbf{n-InAs: $\mathbf{1} \boldsymbol{\times} \mathbf{10^{17}} \, \boldsymbol{\mathrm{cm^{-3}}}$}};
		\fill [cobalt] (0,4)--(10,4)--(10,5.5)--(0,5.5)--(0,4);
		\node[font=\fontsize{10}{60}\selectfont,white] at (5,5.1) {\textbf{BR ($\mathbf{0.5} \, \textbf{\textmu m}$)}};
		\node[font=\fontsize{10}{60}\selectfont,white] at (5,4.35) {\textbf{n-InAs: $\mathbf{1} \boldsymbol{\times} \mathbf{10^{20}} \, \boldsymbol{\mathrm{cm^{-3}}}$}};
		\fill [airforceblue] (0,0.5)--(10,0.5)--(10,4)--(0,4)--(0,0.5);
		\node[font=\fontsize{10}{60}\selectfont] at (5,2.625) {\textbf{Substrate ($\mathbf{500} \, \textbf{\textmu m}$)}};
		\node[font=\fontsize{10}{60}\selectfont] at (5,1.875) {\textbf{n-InAs: $\mathbf{2} \boldsymbol{\times} \mathbf{10^{18}} \, \boldsymbol{\mathrm{cm^{-3}}}$}};
		\fill [yellow] (0,0.5)--(10,0.5)--(10,0)--(0,0)--(0,0.5);
		\draw (0,0.5)--(10,0.5)--(10,0)--(0,0)--(0,0.5);
	
	\end{tikzpicture}
	\caption{Illustration of the InAs-based NFTPV system for which we design an optimized radiator. }\label{fig2}
\end{figure}

Using a one-dimensional fluctuational electrodynamics model from Ref.~\citenum{HeatSlabs}, we attempt to improve NFTPV system performance via optimization of radiator material. Our goal is to determine whether an InAs radiator outperforms a conventional doped Si radiator in an InAs-based NFTPV system. We consider an InAs PV cell designed specifically for high efficiency NFTPV \cite{Forcade_2022}, which was designed assuming a silicon radiator, and compare the performance of a NFTPV system using a silicon or InAs radiator. The PV device modeled in this work consists of a $\text{p-i-n-n}^+$ InAs structure, depicted in Fig.~\ref{fig2}, with the $\text{n}^+$ layer effectively forming a reflector for subgap radiation. The layers in this $\text{p-i-n-n}^+$ structure are designated as the front surface field (FSF), emitter, base, and back reflector (BR), respectively. Note that the high doping level in the BR layer leads to $\omega_\mathrm{p}>E_\mathrm{g}$. As described in Section \ref{subsec:dielectric_function}, the ionized impurity model is employed only for $\omega>\omega_\mathrm{p}$, so the BR layer uses the Drude model and thus has an overestimation of absorptivity. In all simulations, the PV cell is assumed to operate at 300 K. We assess the performance of NFTPV systems using two key metrics: the useful transferred power $P$ and the spectral efficiency $\eta$ \cite{St-Gelais_2017},

\begin{equation}
P=\int^\infty_{\frac{E_\mathrm{g}}{\hbar}}\frac{E_\mathrm{g}}{\hbar\omega}Q(\omega)\mathrm{d}\omega,
\end{equation}\label{eq:Useful_power}

\begin{equation}
\eta=\frac{P}{\int^\infty_0 Q(\omega)\mathrm{d}\omega},
\end{equation}\label{eq:spec_eff}

\noindent where $E_\mathrm{g}$ is the bandgap energy and $Q(\omega)$ is the spectral heat flux. $P$ is an upper limit to useful power, where we assume that all energy transfer with $\omega>E_\mathrm{g}/\hbar$ generates electron-hole pairs, and that they each give $E_\mathrm{g}$ of useful energy, as in Ref.~\citenum{St-Gelais_2017}. The spectral efficiency considers the inefficiencies from above-gap heat transfer, where the PV cell cannot use all of the energy, as well as subgap heat transfers that do not produce electron-hole pairs but still heat the PV cell.\par

The radiative heat flux absorbed by the PV device depends on several factors, including the radiator doping level and thickness, gap, and temperature. Therefore, prior to comparing the performance of the two radiator materials, we must optimize the radiator parameters for various fixed gaps and temperatures.\par

\subsubsection{Optimization of the Radiator Parameters}
\label{subsubsec:parameter_optimization}

Before comparing radiator materials, we must first select appropriate thicknesses and doping levels. To ensure a fair comparison, we optimize these parameters for each radiator material. We choose four representative test conditions (700 and 1000 K radiator temperatures with 30 and 100 nm gaps) and optimize the radiator thickness and doping level at each test condition. Our goal is to select a radiator design that can perform well at a variety of temperatures and gaps. We first identify optimal designs for each test condition, and then select the design with the best average performance across the four test conditions. \par

We consider both finite and semi-infinite thickness radiators separately. We analyze the performance of semi-infinite thickness radiators to study the behavior of bulk materials. We simulate semi-infinite thickness radiators by generating an output medium behind the radiator made of the same material, neglecting any transmission through the layer. This semi-infinite medium removes any spurious Fabry-Pérot interference effect that can arise in the case of a finite stack. Radiators can effectively behave as semi-infinite at achievable thicknesses as low as \SI{100}{\micro\meter} depending on the material and doping level, and this approach enables us to compute the performance of such bulk radiators much faster than if we considered large finite-thickness radiators.\par

We optimize for the quantity $P\cdot\eta$ since we want large power densities with high efficiency. Optimizing either $P$ or $\eta$ on its own does not capture these effects. Although optimizing $P$ is essential, it is equally important to optimize $\eta$, as poor spectral efficiency leads to significant parasitic subgap radiation, heating the cell and increasing cooling requirements. \par 

Our optimizer relies on the Nelder-Mead method implemented in the Optim package in Julia. Initial guesses are generated using a coarse $5\times5$ grid for the doping level and radiator thickness. After finding optimal finite and semi-infinite designs for each material under each test condition, we normalize the $P\cdot\eta$ value by dividing it by its maximum across all radiator designs for each test condition. We then compute the average normalized $P\cdot\eta$ value across all test conditions and select the design with the largest average. The selected optimal finite and semi-infinite designs are reported in Table \ref{Table:Optimized_designs}. In Table \ref{Table:Optimized_designs}, we note that, for Si, a semi-infinite thickness radiator offers better performance, while for InAs, a finite thickness radiator is more effective.\par

\begin{table*}[t] 
\caption{Optimized finite and semi-infinite NFTPV radiator designs with their performance at each test condition.}\label{Table:Optimized_designs}
\centering
\begin{tabular}{c c c c c c c c}
  \hline
  Radiator & Radiator & Thickness & Doping & \multicolumn{4}{c}{$P\cdot\eta$ [\SI{}{\kilo\watt\per\square\centi\meter}]} \\ 
  \cline{5-8}
  Design & Material & [\SI{}{\micro\meter}] & [\SI{}{\per\cubic\centi\meter}] & \multicolumn{2}{c}{\SI{700}{\kelvin}} & \multicolumn{2}{c}{\SI{1000}{\kelvin}} \\
  \cline{5-8}
   &  &  &  & \SI{30}{\nano\meter} & \SI{100}{\nano\meter} & \SI{30}{\nano\meter} & \SI{100}{\nano\meter} \\
  \hline
  Bulk InAs & n-doped InAs & $\infty$ & $4.6\times10^{18}$ & 4.89 & 3.13 & 68.5 & 42.7 \\
  Finite InAs & n-doped InAs & 6.5 & $2\times10^{18}$ & 5.15 & 3.56 & 68.8 & 43.6 \\
  Bulk Si & p-doped Si & $\infty$ & $1\times10^{17}$ & 4.51 & 2.92 & 62.1 & 43.3 \\
  Finite Si & p-doped Si & 100 & $1\times10^{20}$ & 2.94 & 2.45 & 52.4 & 40.3 \\
  \hline
\end{tabular}
\end{table*}

\subsubsection{InAs vs Si}
\label{subsubsec:InAs_vs_Si}

\begin{figure}[!htb]
	\begin{subfigure}{\columnwidth}
		\captionsetup{justification=raggedright, singlelinecheck=false, position=top} 
		\caption{}\label{fig_3:sub1}
		\includegraphics[width=\columnwidth]{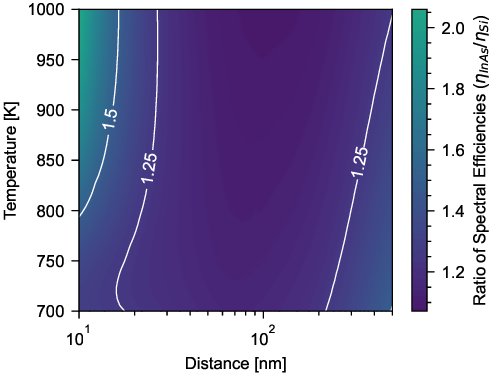}
		\vspace{-5 mm}
	\end{subfigure}
	\begin{subfigure}{\columnwidth}
		\captionsetup{justification=raggedright, singlelinecheck=false, position=top} 
		\caption{}\label{fig_3:sub2}
		\includegraphics[width=\columnwidth]{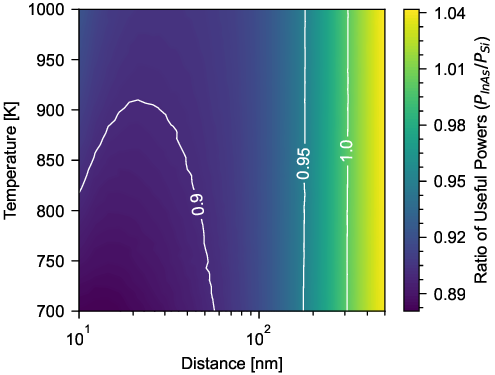}
	\end{subfigure}
	\caption{Bulk InAs vs Bulk Si: (\subref{fig_3:sub1}) Spectral efficiency ratio and (\subref{fig_3:sub2}) useful power ratio. For bulk radiators, InAs offers greater spectral efficiency at the expense of a relative reduction in useful power of over 11\% at small gaps.}\label{fig3}
\end{figure}

We first compare, in Fig.~\ref{fig3}, the performance of the two semi-infinite radiators, one InAs and the other Si, for which the doping level was optimized (Bulk InAs and Bulk Si designs). In the case of a semi-infinite radiator, InAs achieves significantly higher spectral efficiency, particularly at small gaps and elevated temperatures. InAs, however, suffers from more than 11\% relative reduction in useful power at small gaps.\par

\begin{figure}[!htb]
	\begin{subfigure}{\columnwidth}
		\captionsetup{justification=raggedright, singlelinecheck=false, position=top} 
		\caption{}\label{fig_4:sub1}
		\includegraphics[width=\columnwidth]{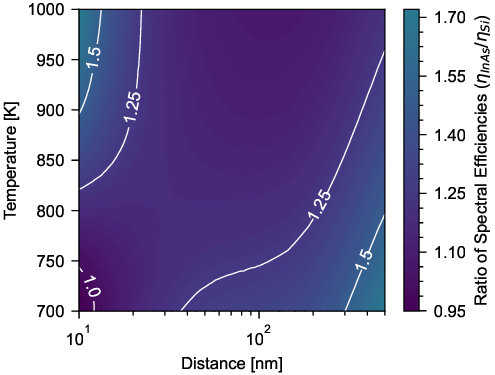}
		\vspace{-5 mm}
	\end{subfigure}
	\begin{subfigure}{\columnwidth}
		\captionsetup{justification=raggedright, singlelinecheck=false, position=top} 
		\caption{}\label{fig_4:sub2}
		\includegraphics[width=\columnwidth]{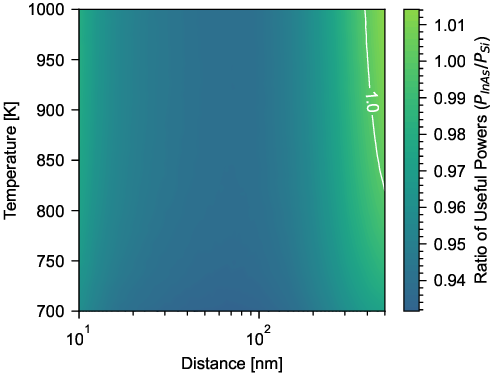}
	\end{subfigure}
	\caption{Best-case InAs vs best-case Si, i.e., Finite InAs vs Bulk Si: (\subref{fig_4:sub1}) Spectral efficiency ratio and (\subref{fig_4:sub2}) useful power ratio. An InAs radiator offers greater spectral efficiency and comparable useful power. }\label{fig4}
\end{figure}

In Table \ref{Table:Optimized_designs}, we note that, for Si, a semi-infinite radiator offers the best performance, while for an InAs radiator, a finite-thickness radiator is more effective. In fact, the Finite InAs radiator improves the average $P\cdot\eta$ by approximately 5\%. We, therefore, compare these two best cases (Finite InAs design vs Bulk Si design) in Fig.~\ref{fig4}. The Finite InAs radiator significantly improves spectral efficiency at larger gaps, while offering a useful power similar to that of a silicon radiator. Conversely, at smaller gaps, the thinner InAs radiator achieves greater useful power than the semi-infinite and is at most only 7\% lower than the Si radiator, though this power comes with a small reduction in the spectral efficiency.\par

\begin{figure}[!htb]
	\begin{subfigure}{\columnwidth}
		\captionsetup{justification=raggedright, singlelinecheck=false, position=top} 
		\caption{}\label{fig_5:sub1}
		\includegraphics[width=\columnwidth]{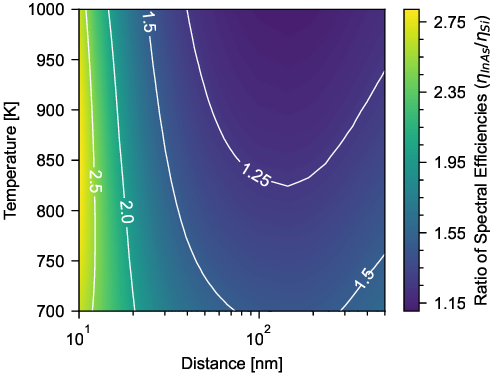}
		\vspace{-5 mm}
	\end{subfigure}
	\begin{subfigure}{\columnwidth}
		\captionsetup{justification=raggedright, singlelinecheck=false, position=top} 
		\caption{}\label{fig_5:sub2}

		\includegraphics[width=\columnwidth]{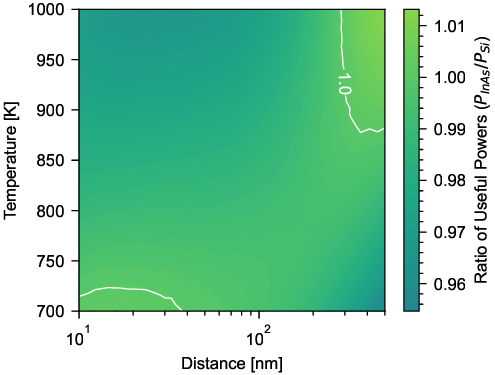}
	\end{subfigure}
	\caption{Finite InAs vs Finite Si: (\subref{fig_5:sub1}) Spectral efficiency ratio and (\subref{fig_5:sub2}) useful power ratio. For thin-film radiators, InAs offers greater spectral efficiency, achieving nearly a threefold enhancement at small distances, with a reduction in useful power of at most 5\%.}\label{fig5}
\end{figure}

Since experimental NFTPV platforms often rely on thin films, we compare, in Fig.~\ref{fig5}, the performance of the two radiator materials using finite-thickness designs ($< \SI{100}{\micro\meter}$). In this case, an InAs radiator significantly improves the spectral efficiency, achieving nearly a threefold enhancement at small distances, while only resulting in a reduction in useful power of at most 5\%. \par

\subsubsection{Underlying Phenomena}
\label{subsubsec:underlying_phenomena}

\begin{figure}[!htb]
	\includegraphics[width=\columnwidth]{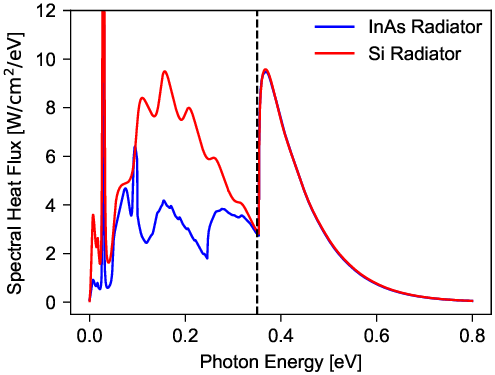}
	\caption{Spectral radiative heat transfer in an InAs-based NFTPV system using an InAs radiator (Finite InAs design, blue) and a Si radiator (Finite Si design, red) at 700 K and a separation of 100 nm.}\label{fig6}
\end{figure}

\begin{figure}[!htb]
	\begin{subfigure}{\columnwidth}
		\captionsetup{justification=raggedright, singlelinecheck=false, position=top} 

		\includegraphics[width=\columnwidth]{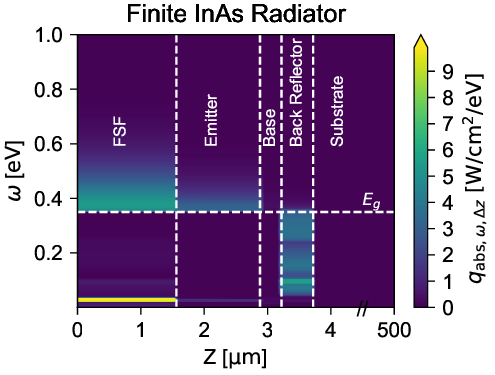}
		\vspace{-68 mm}
		\caption{}\label{fig_7:sub1}
		\vspace{61 mm}
	\end{subfigure}
	\begin{subfigure}{\columnwidth}
		\captionsetup{justification=raggedright, singlelinecheck=false, position=top} 
		
		\includegraphics[width=\columnwidth]{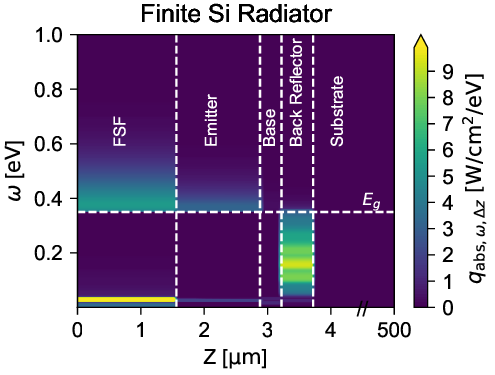}
		\vspace{-68 mm}
		\caption{}\label{fig_7:sub2}
		\vspace{61 mm}
	\end{subfigure}
	\caption{Spectral-spatial absorption distribution, $q_{\mathrm{abs},\omega,\Delta z}$, within the InAs cell for a separation of 100 nm and a radiator temperature of 700 K using (\subref{fig_7:sub1}) an InAs radiator (Finite InAs design) and (\subref{fig_7:sub2}) a Si radiator (Finite Si design).}\label{fig7}
\end{figure}

To study the origin of the spectral efficiency enhancement observed with an InAs radiator, Fig.~\ref{fig6} compares the spectral heat flux of both radiator materials. Compared to the Si radiator, subgap heat absorption is substantially reduced with the InAs radiator, without significantly affecting the above-bandgap heat absorption. Fig.~\ref{fig7} shows the spatially resolved spectral absorption in the PV cell, clarifying that the InAs radiator significantly reduces the below-bandgap absorption in both the front-surface p-region, and in the $\text{n}^+$ layer that acts as a back reflector. \par

\begin{figure}[!htb]
	\includegraphics[width=\columnwidth]{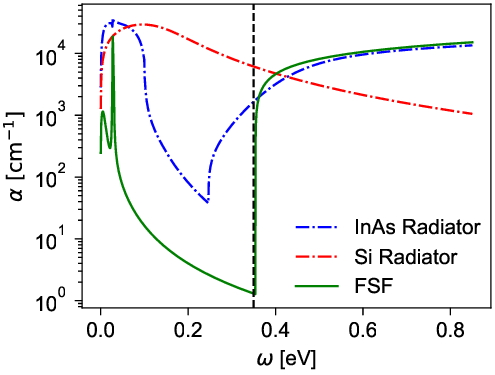}
	\caption{Absorption coefficient of the FSF layer of the InAs PV cell at 300 K (green), compared with the absorption coefficient of an InAs radiator (Finite InAs design, blue) and a Si radiator (Finite Si design, red), both at 700 K.}\label{fig8}
\end{figure}

To study how the optical properties of InAs results in such improvement, Fig.~\ref{fig8} contrasts the absorption coefficient of the FSF to the InAs radiator (i.e., doping of the Finite InAs design), and the Si radiator (i.e., doping of the Finite Si design). This comparison shows that the absorption coefficient of InAs is much lower than that of Si at energies just below the bandgap. According to the Kirchoff law, under isothermal conditions, the emissivity of a material is proportional to its absorption coefficient. Therefore, the absorption coefficient of the radiator plays a crucial role in determining the radiative heat flux. In this case, the lower subgap absorption coefficient of the InAs radiator could explain the reduction in subgap absorption in the BR layer.\par

\begin{figure}[!htb]
	\begin{subfigure}{\columnwidth}
		\captionsetup{justification=raggedright, singlelinecheck=false, position=top} 
		
		\includegraphics[width=\columnwidth]{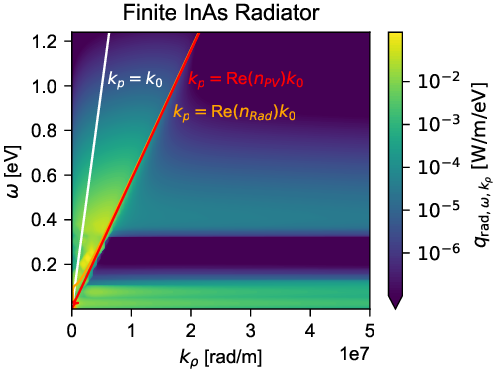}
		\vspace{-68 mm}
		\caption{}\label{fig_9:sub1}
		\vspace{61 mm}
	\end{subfigure}
	\begin{subfigure}{\columnwidth}
		\captionsetup{justification=raggedright, singlelinecheck=false, position=top} 
		
		\includegraphics[width=\columnwidth]{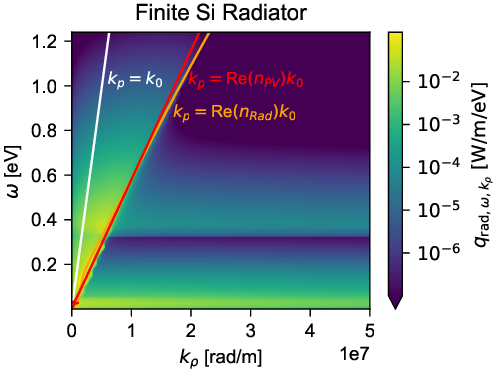}
		\vspace{-68 mm}
		\caption{}\label{fig_9:sub2}
		\vspace{61 mm}
	\end{subfigure}
	\caption{Radiative heat flux, $q_{\mathrm{rad},\omega,k_{\rho}}$, per unit angular frequency, $\omega$, and parallel wavevector, $k_\rho$, at a 10 nm separation and a radiator temperature of 700 K for (\subref{fig_9:sub1}) an InAs radiator (Finite InAs design) and (\subref{fig_9:sub2}) a Si radiator (Finite Si design). For both materials, the above-bandgap radiative heat flux is primarily influenced by frustrated modes ($k_0<k_\rho<k_0 \mathrm{min}\left(\mathrm{Re}(n_\mathrm{radiator}),\mathrm{Re}(n_\mathrm{PV})\right)$), while propagating modes ($k_\rho<k_0$) have modest contributions. For Si, the below-bandgap radiative heat flux is significantly dominated by surface polariton resonances ($k_\rho>k_0 \mathrm{min}\left(\mathrm{Re}(n_\mathrm{radiator}),\mathrm{Re}(n_\mathrm{PV})\right)$).}\label{fig9}
\end{figure}

We attribute the substantial below-bandgap absorption observed with the Si radiator to a surface polariton resonance. As shown in Fig.~\ref{fig9}, for both radiator materials, at a 10 nm separation, the above-bandgap heat transfer is primarily influenced by frustrated modes occurring at a parallel wavevector $k_\rho$ exceeding the vacuum wavevector $k_0$ but lower than the material light lines (given by $\mathrm{Re}(n) k_0$ where $n$ is the refractive index of the material) \cite{Desutter_2019}. These results also demonstrate that, for Si, the below-bandgap radiative heat flux is dominated by a resonance at $\omega=\SI{0.0255}{\eV}$. This resonance originates from surface polariton resonances in the Si radiator, since its parallel wavevector exceeds both material light lines. This below-bandgap surface resonance is anticipated since silicon has a plasma resonance below bandgap ($\omega_\mathrm{p}=\SI{0.198}{\eV}$ for the Finite Si design \cite{Basu_2010}). Conversely, InAs is a polar material with a resonance around 0.028 eV, which is slightly noticeable in Fig.~\ref{fig_9:sub1}. However, this resonance is much narrower than the plasmonic resonance in Si, leading to less below-bandgap emission. Consequently, the presence of a slowly decaying below-bandgap surface polariton resonance in Si explains why an InAs radiator significantly enhances spectral efficiency. \par

\section{Conclusions}
\label{sec:conclusions}

We have demonstrated that the spectral efficiency of NFTPV systems can be greatly enhanced when both the PV cell and radiator are made of InAs. In such a case, spectral coupling results in a significant improvement in spectral efficiency compared to a conventional doped Si radiator, achieving nearly a threefold enhancement at small distances. This improvement reduces the parasitic heating of the PV cell and increases its overall efficiency while not significantly affecting the power density. Additionally, utilizing InAs for both the PV cell and the radiator could be advantageous for monolithic fabrication in a single crystal growth process for both the radiator and the PV cell. During our analysis, we also uncovered an overestimation of free carrier absorption in InAs dielectric function models. We implement a corrective model and demonstrate that it accurately represents the absorption at moderate doping levels but could be further refined for improved accuracy at higher doping levels. \par

\appendix
\section{Kramers-Kronig Relations in InAs Dielectric Model}
\label{Appendix_A}

In Ref.~\citenum{Milovich_2020}, the real part of the interband refractive index $n_{\mathrm{IB}}^\mathrm{\prime}$ is calculated from the imaginary part of the interband refractive index $n_{\mathrm{IB}}^\mathrm{\prime\prime}$, using the Kramers-Kronig relations. Similarly, the contribution from Eq.~\ref{eq:Baltz_model} to the real part of the permittivity $\varepsilon^\mathrm{\prime}$, can be determined using the Kramers-Kronig relations. Unfortunately, the Kramers-Kronig relations can be computationally expensive. Therefore, to avoid a Kramers-Kronig calculation, we instead approximate $n_{\mathrm{IB}}^\mathrm{\prime}=\sqrt{\varepsilon_\infty}$, following the approach of Ref.~\citenum{Forcade_2022}, and employ the Drude model to describe the real part of the permittivity due to free carrier absorption. We find that this approximation results in an underestimation of both the useful transferred power and the spectral efficiency of approximately 1\% and 4\%, relative, respectively.\par

Note that NFTPV simulations only require optical constants up to approximately 1.5 eV, although the Kramers-Kronig relations involve integration over the entire frequency range. In our dielectric model, the interband absorption model, taken from Ref.~\citenum{Anderson_1980}, is based on $k\cdot p$ theory and is therefore only valid near the absorption edge. In this case, the light-hole absorption contribution obeys $\alpha_\mathrm{lh}(\omega)\sim\omega$ as $\omega \to \infty$, leading to a non-convergent Kramers-Kronig relation. This limit is not physical, and the $\alpha_\mathrm{lh}$ term can be smoothly suppressed at energies above 1 eV with only a few percent change in the Kramers-Kronig derived interband refractive index $n_{\mathrm{IB}}^\mathrm{\prime}$ depending on the exact form of the suppression function. We consider such changes to be within the expected errors for the modeled optical constants. Also note that we evaluate the interband absorption coefficient $\alpha_\mathrm{IB}$ by considering only the contributions from the light-hole and heavy-hole bands, as the contributions from remote bands are accounted for in $\varepsilon_\infty$. Absorption from the split-off hole band begins at approximately 0.75 eV, and above this energy, $\alpha_\mathrm{IB}$ is already underestimated.\par

\section*{Acknowledgments}

The authors thank Mathieu Francoeur from the University of Utah and Daniel Milovich from the Universidad Politécnica de Madrid for useful discussions on the dielectric model of InAs. The authors also thank Mauro Antezza from the Institut Universitaire de France for stimulating discussions on the Kramers-Kronig relations. The authors also extend their gratitude to University of Ottawa colleague Gavin Forcade for discussions on permittivity and heat transfer calculation algorithms. This work was funded by the National Science and Engineering Research Council of Canada (NSERC) through the CREATE TOP-SET program (No. 497981) and the CGS-D program. 

\bibliographystyle{elsarticle-num-arXiv} 
\bibliography{refs}

\end{document}